\documentclass[letter]{aa} 

\newcommand{\msun}{\mbox{M$_\odot$}}

\usepackage{graphicx}
\usepackage{txfonts}
\usepackage{natbib}
\bibpunct{(}{)}{;}{a}{}{,} 
\begin{document}
   \title{The mass function of dense molecular cores \\and the origin
  of the IMF}


   \author{J. Alves
          \inst{1}
          \and
          M. Lombardi\inst{2}\fnmsep\thanks{University of Milan, Department of Physics, via Celoria 16,  I-20133 Milan, Italy}
          \and
          C. J. Lada\inst{3}
          }

   \offprints{J. Alves}

   \institute{Calar Alto Observatory -- Centro Astron\'omico Hispano
  Alem\'an, C/Jes\'us Durb\'an Rem\'on 2-2, 04004 Almeria, Spain\\
              \email{jalves@caha.es}
         \and
             European Southern Observatory, Karl-Schwarzschild-Str. 2
    85748 Garching, Germany\\
             \email{mlombard@eso.org}
         \and
Harvard-Smithsonian Center for Astrophysics, Mail Stop 72, 60 Garden
  Street, Cambridge, MA 02138\\
             \email{clada@cfa.harvard.edu}
             }

   \date{Received August 2006; accepted... }

 
  \abstract
  {Stars form in the cold dense cores of interstellar molecular clouds
    and the detailed knowledge of the spectrum of masses of such cores
    is clearly a key for the understanding of the origin of the
    IMF. To date, observations have presented somewhat contradictory
    evidence relating to this issue.}
  {In this paper we propose to derive the mass function of a complete
    sample of dense molecular cores in a single cloud employing a
    robust method that uses uses extinction of background starlight to
    measure core masses and enables the reliable extension of such
    measurements to lower masses than previously possible.}
  {We use a map of near-infrared extinction in the nearby Pipe dark
    cloud to identify the population of dense cores in the cloud and
    measure their masses.}
  {We identify 159 dense cores and construct the mass function for
    this population. We present the first robust evidence for a
    departure from a single power-law form in the mass function of a
    population of cores and find that this mass function is
    surprisingly similar in shape to the stellar IMF but scaled to a
    higher mass by a factor of about 3.  This suggests that the
    distribution of stellar birth masses (IMF) is the direct product
    of the dense core mass function and a uniform star formation
    efficiency of 30\%$\pm$10\%, and that the stellar IMF may already
    be fixed during or before the earliest stages of core
    evolution. These results are consistent with previous dust
      continuum studies which suggested that the IMF directly
      originates from the core mass function.  The typical density of
    $\sim$10$^4$ cm$^{-3}$ measured for the dense cores in this cloud
    suggests that the mass scale that characterizes the dense core
    mass function may be the result of a simple process of thermal
    (Jeans) fragmentation.}
   {}

   \keywords{ISM: structure, dust, extinction --- Stars: formation ---
   Stars: mass function}

   \maketitle
%

\section{Introduction}

The stellar initial mass function (IMF) is one of the most fundamental
distributions in astrophysics.  Detailed knowledge of its functional
form and how this form varies in space and time is necessary to
predict and understand the evolution of stellar systems, from 
clusters to galaxies.  
\noindent
Nevertheless, the origin of the stellar IMF remains one of the major
unsolved problems of modern astrophysics.  

\begin{figure*}
  \begin{center}
    \includegraphics[scale=0.9]{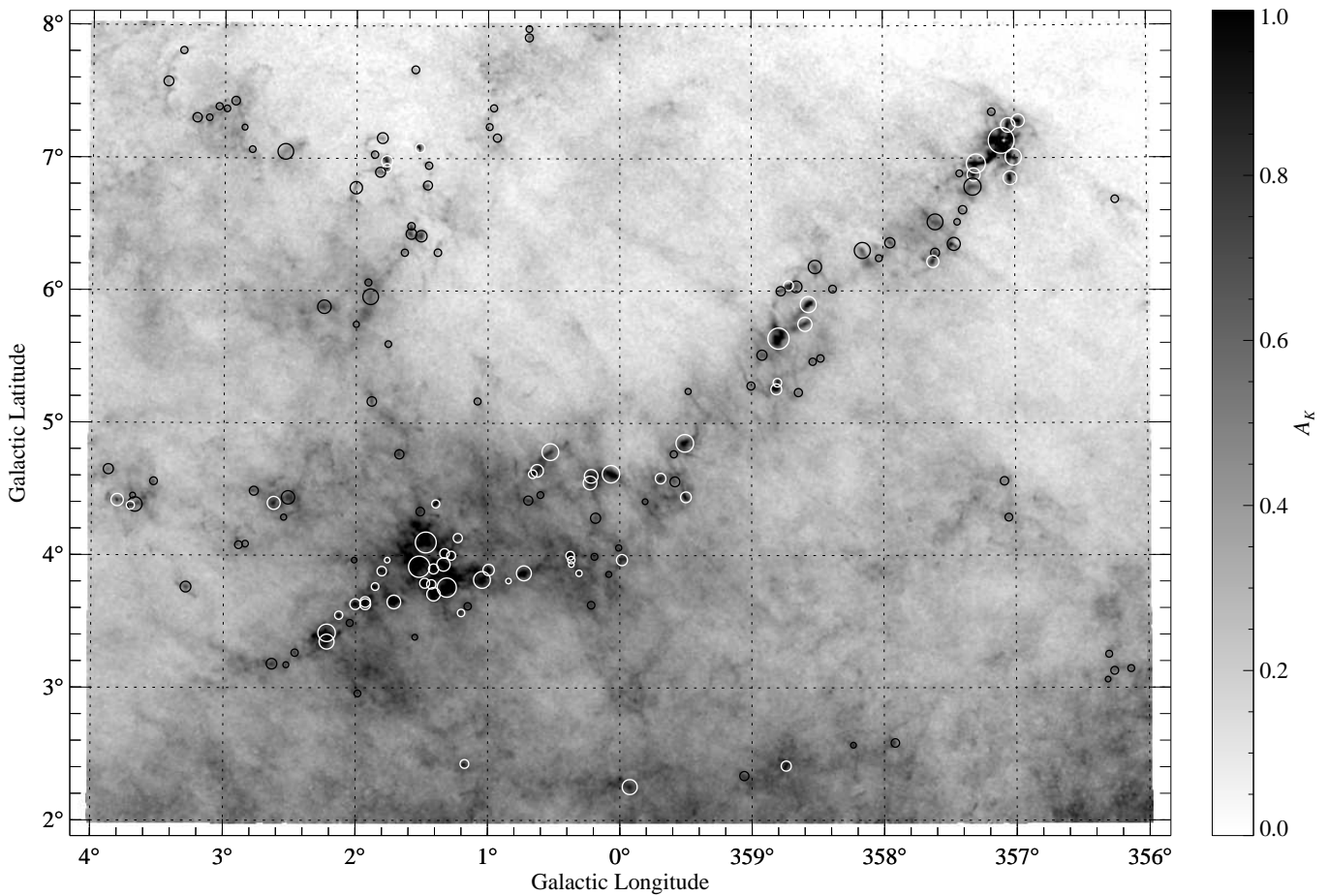}    
    \caption{Dust extinction map of the Pipe nebula molecular complex
      from \citet{Lombardi2006}. This map was constructed from
      near-infrared observations of about 4 million stars in the
      background of the complex. Approximately 160 individual cores
      are identified within the cloud and are marked by an open circle
      proportional to the core radius.  Most of these cores appear as
      distinct, well separated entities.\label{fig:1}}
  \end{center}
\end{figure*}

Stars form in interstellar molecular clouds and the detailed knowledge
of the spectrum of structure in such clouds is likely a key for the
understanding of the origin of the IMF.  What is the relation between
cloud structure and the IMF?  To date, observations have presented
somewhat contradictory evidence relating to this issue.  Early CO
observations suggested that the mass spectrum of molecular cloud
structure is well described by a single power-law
\citep{Blitz1993,Kramer1998,Heithausen1998}, specifically,
d$N(m)/$d$log\,m \sim m^{-0.7}$ for structures (clumps) with a wide
range of masses that apparently extends from a fraction a solar mass
to 1000 \msun.  This functional form differs from the form of the
stellar IMF in two fundamental ways.

First, the stellar IMF cannot be described by a single power-law
(scale-free) function.  The general consensus of observational studies
is that the stellar IMF is best described by a series of power-law
segments that overall is qualitatively similar in shape to a
log-normal function \citep{Miller1979,Kroupa2002,Muench2001}.  In
particular, the IMF is characterized by a broad peak which spans
roughly the interval of 0.1--0.6 \msun\ and falls off rapidly on each
side of the peak.  Between 0.6 and 10 \msun\ the IMF decreases in a
power-law fashion similar to that originally derived by
\citet{Salpeter1955}.  The fact that the IMF displays a peak and is
not scale-free is of great significance for star formation since it
indicates that there is a characteristic mass resulting from the star
formation process of around 0.2--0.3 \msun\
\citep[e.g.,][]{Larson2005}.

Second, above about a solar mass the slope of the stellar IMF is
steeper (i.e., d$ N(m)/$d$log\,m \sim m^{-1.35}$) than the cloud mass
spectrum.  This latter difference implies a significant physical
difference in the two distributions.  For the stellar IMF, the bulk of
the (stellar) mass is tied up in low mass objects while for clouds the
bulk of the mass is tied up in the most massive objects.  A
consequence of this difference is that in order to produce the stellar
IMF from the cloud mass spectrum, a transformation must take place
during the process of star formation.  It has been suggested that
outflows generated during the protostellar stages of star formation
provide a natural feedback to collapse/infall limiting the final mass
of a protostellar object \citep{Shu1987} and perhaps providing the
mechanism for transforming the form of the cloud mass spectrum into
the form of the stellar mass spectrum \citep{Lada2003}.  However, it
is not clear how such a process could produce a peak, or a
characteristic scale for stellar masses.  Moreover, the comparison of
the CO core mass function and the IMF may not be relevant since CO
does not trace the dense component of the molecular gas within which
stars actually form \citep{Lada1992a}.

Indeed, a different picture appears to emerge when observations of
dense gas are considered.  Typically only about 10\% or less of the
mass of a star-forming molecular cloud is in the form of dense (i.e.,
$n(\mathrm{H2}) \sim 10^4 \mbox{cm}^{-3}$) gas and this gas appears to
be organized into discrete cores within which stars form.
\citet{Tachihara2002} and \citet{Onishi2002}, using C$^{18}$O and
H$^{13}$CO$^+$ as tracers of dense gas, suggest that the stellar and
core mass distributions are similar.  Recent observations of dust
continuum emission originating from such dense cores has enabled the
construction of the dense core mass function (DCMF) in a number of
nearby molecular cloud complexes.  For cores with masses in excess of
$\sim$0.5 \msun\ the derived mass spectra appear to be described by a
single power-law, similar to the CO cores but with a relatively steep
slope ($-1.1$ to $-1.6$, in log mass units) similar to the that of the
stellar IMF
\citep{Motte1998,Testi1998,Johnstone2000,Johnstone2001,Motte2001,Beuther2004,Stanke2006}.
Moreover, in one example, the $\rho$ Ophiuchi cloud, the core mass
spectrum exhibited possible evidence of a flattening or break near a
mass of about 0.5 \msun, also similar to the stellar IMF
 \citep{Motte1998}.  In another cloud, NGC 1333, measurements of dust
emission produced a core mass spectrum between 0.1 and 0.5 \msun\ with
a slope of approximately $-$0.4 \citep{Sandell2001}, similar to that
universally derived for less-dense gas traced by CO emission in other
clouds but also consistent with the apparent break and assumed
flattening below 0.5 \msun\ of the DCMF of the Ophiuchi dark cloud
mentioned above.  The observed similarity between the slopes of the
DCMF derived from millimeter-wave dust emission studies and the slope
of the IMF above 0.5 \msun\ has been taken as evidence that the
individual dense cores are the direct precursors of new stars and
moreover that the stellar IMF is completely specified by the
fragmentation process in molecular clouds.  In addition, the possible
flattening of the DCMF near 0.5 \msun\ implies a high star formation
efficiency (SFE) for dense gas (about 100\%).  In this case there is
no need for a mass transformation from the DCMF to the IMF.  The
characteristic scale of stellar mass demanded by the stellar IMF is
set by the fundamental physics of cloud fragmentation
\citep{Shu2004,Larson2005}.  Which picture is correct?

\begin{figure}
  \begin{center}
    \includegraphics[scale=1.07]{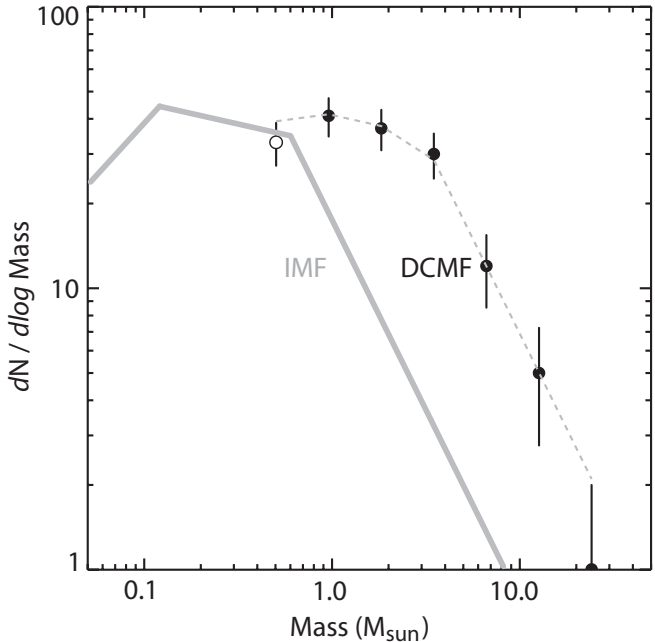}
    \caption{Mass function of dense molecular cores plotted as filled
      circles with error bars. The grey line is the stellar IMF for
      the Trapezium cluster \citep{Muench2002}. The dashed grey line
      represents the stellar IMF in binned form matching the
      resolution of the data and shifted to higher masses by about a
      factor of 4. The dense core mass function is similar in shape to
      the stellar IMF function, apart from a uniform star formation
      efficiency factor. \label{fig:2}}
  \end{center}
\end{figure}

In order to bring new insight to this issue we used an independent
method of identifying and measuring the masses of dense cores.  This
method uses precise infrared measurements of dust extinction toward
stars background to a molecular cloud (the \textsc{Nice(r)} method,
\cite{Lada1994,Alves1998,Alves2001,Lombardi2001}).  Such measurements
are free from many of the complications and systematic uncertainties
that plague molecular-line or dust emission data and thus enable
robust maps of cloud structure to be constructed.  We used data from
the recent wide-field extinction map of the Pipe Nebula constructed by
\citet{Lombardi2006}, hereafter LAL06, using 2MASS data.  The Pipe
nebula is a virtually unstudied nearby molecular cloud complex
(\cite{Onishi1999}, LAL06), at a distance of about $130\, \mbox{pc}$
and with a total mass of $\sim 10^4 \mbox{ M}_\odot$.


\section{Observations}

The details of the extinction study are described in LAL06. Briefly,
this molecular complex was selected because 1) this is {\bf one of}
the closest to Earth complex of this size and mass, 2) it is
particularly well positioned along a relatively clean line of sight to
the rich star field of the Galactic bulge, which given the close
distance of the Pipe nebula allowed us to achieve spatial resolutions
of $\sim 0.03\, \mbox{pc}$, or about 3 times smaller the typical dense
core size, and 3) it exhibits very low levels of star formation
suggesting that its dense cores likely represent a fair sample of the
initial conditions of star formation. LAL06 applied a 3-band ($1.25\
\mu\mbox{m}$, $1.65\ \mu\mbox{m}$, $2.2\ \mu\mbox{m}$) optimized
version of this method, the \textsc{Nicer} method
\citep{Lombardi2001}, to about 4 million stars background to the Pipe
nebula complex to construct a $6^\circ \times 8^\circ$ dust column
density map of this complex, presented in Figure~\ref{fig:1}.
Because of the high dynamic range in column density achieved by this
map (3$\sigma\sim$ 0.5 $<$ A$_V$ $<$ 24 mag or $10^{21} < N <
5\times10^{22}$ cm$^{-2}$), cores are easily visually identified as
high contrast peaks embedded on rather smooth but variable background
(see Figure~\ref{fig:1}). Unfortunately, because of the high dynamic
range and variable background, traditional source extraction
algorithms based on thresholding fail to identify these objects in a
coherent way.  An alternative approach is to extract cores based on
their sizes, using a multi-scale algorithm. For this study we used a
algorithm developed by Vandame (2006, private comm.), which, in brief,
uses the wavelet transform of the image to first identify and then
reconstruct the dense cores\footnote{\emph{Object identification in
    wavelet space:} for a given scale $i$, structures are isolated
  with classical thresholding at $3\sigma_{i}$ with $\sigma_{i}$ being
  the noise amplitude at scale $i$.  A structure at scale $i$ is
  connected with a structure at scale $i+1$ if its local maxima drops
  in the structure at scale $i+1$. The size scales considered were
  2$^\prime$, 4$^\prime$, and 8$^\prime$ (0.08 pc, 0.15 pc, and 0.30
  pc). One then builds a 3D distribution of significant structures (x,
  y, and $i$). The algorithm developed by Vandame (2006, private
  comm.)  offers rules that split the 3D distribution into independent
  trees corresponding to one core and its corresponding hierarchical
  details. \emph{Object reconstruction:} This same algorithm performs
  a complex iterative reconstruction of the cores following the trees
  defined in the previous step. The final ``cores only'' image is
  validated by subtracting it from the extinction map, effectively
  creating a smooth image of the variable background.}.  This step
defines the projected core boundaries.

Masses are derived by integrating the extinction map over the area of
each core and multiplying by the standard gas-to-dust ratio.  The
final Pipe core sample has 159 objects with effective diameters
between 0.1 and 0.4 pc (median size is 0.18 pc) and peak extinctions
that range from 3.0 to 24.3 visual magnitudes (mean extinction is 8.4
magnitudes).  The derived core masses range between 0.5 to $28 \mbox{
  M}_\odot$. The assessment of sample completeness is non trivial
because of the variable background.  Nevertheless, the completeness
should not be dominated by confusion but sensitivity, as the mean
separation between cores, even in the clustered regions, is well above
the resolution of the map. We estimate, conservatively, that the
sample is 90\% complete at about 1 \msun. The mean diameter of a 1 \msun\
core is $\sim$0.2 pc, i.e., about six times the resolution of the map.

\section{Results: The Dense Core Mass Function (DCMF)}

The dense core mass function we derive from the above observations is
presented in Figure~\ref{fig:2}\footnote{The full dataset is available
  in electronic form at the CDS via anonymous ftp to
  cdsarc.u-strasbg.fr (130.79.128.5) or via
  http://cdsweb.u-strasbg.fr/cgi-bin/qcat?J/A+A/}. For comparison we
plot the Trapezium cluster stellar IMF as a grey solid line
\citep{Muench2002}. This IMF consists of 3 power law segments with
breaks and 0.6\msun\ and 0.12\msun\.  We find that the DCMF for the
Pipe Nebula is surprisingly similar in shape to the stellar
IMF. Specifically, both distributions are characterized by a
Salpeter-like power-law \citep{Salpeter1955} that rises with
decreasing mass until reaching a distinct break point, this is then
followed by a broad peak.  Although similar in shape, the stellar and
core mass functions are characterized by decidedly different mass
scales.  The grey dashed line in Figure~\ref{fig:2} is not a fit to
the data but simply the stellar IMF in binned form matched to the
resolution of the data, and shifted by a factor of 4 to the higher
masses.  The break from the Salpeter-like slope seems to occur between
2 and 3$\mbox{ M}_\odot$ for the cloud cores instead of the $0.6
\mbox{ M}_\odot$ for the stellar case \citep{Muench2002}.

The DCMF in Figure~\ref{fig:2} likely suffers from two sources of
uncertainty: 1) the individual core masses are likely upper limits to
the true values since we made no corrections for the local extended
background on which most cores are embedded, and 2) the particular
binning we chose may not produce the most accurate representation of
the underlying mass distribution.  To address point 1) we estimated a
lower limit to the true core masses (and the DCMF) by subtracting from
each core its local background. We then constructed a background
subtracted DCMF and found that \emph {the form of the mass
  distribution was maintained} although the new distribution was
shifted towards the low masses. To address point 2) we compute a
better functional representation for the DCMF using a probability
density estimator that does not require data binning. We used a
gaussian kernel estimator with a window width of 0.15 in units of log
mass \citep{Silverman1986}.

\begin{figure}
  \begin{center}
    \includegraphics[scale=1.07]{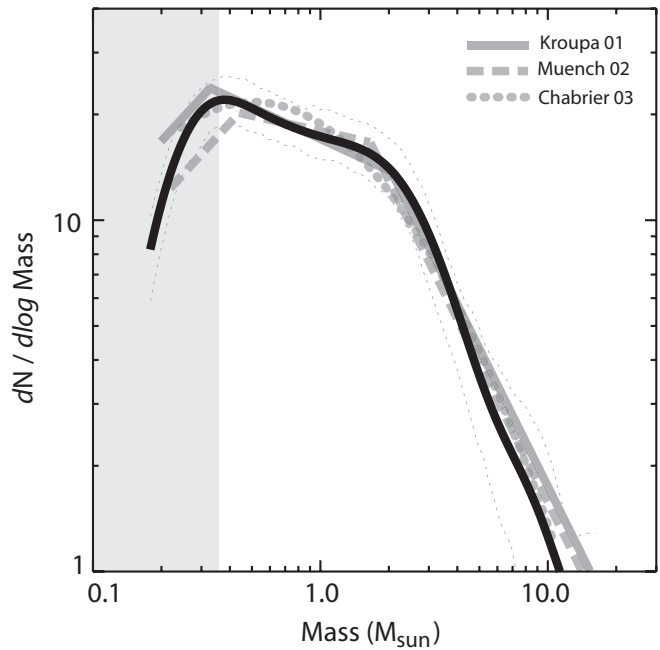}
    \caption{Probability density function of core masses (black) for
      the background subtracted sample. The grey region indicates
      sample incompleteness while the two thin dashed line indicates
      the 90\% confidence limits. The core mass distribution seems to
      be characterized by two power-laws and a well defined break
      point at around 2 \msun. Also plotted are the field star IMFs of
      \citet{Kroupa2001} (solid grey) and \citet{Chabrier2003} (dotted
      grey), and the \citet{Muench2002} (dashed grey) IMF all scaled
      up by a factor of $\sim$ 3 in mass. \label{fig:3}}
  \end{center}
\end{figure}

We present in Figure~\ref{fig:3} the probability density function for
the core masses (background subtracted). As already suggested in
Figure 2, the core mass distribution seems to be characterized by two
power-laws and a well defined break point at $\sim$ 2 \msun. For
comparison we also present the field star IMF determined by
\citet{Kroupa2001} (solid grey) and \citet{Chabrier2003} IMF (dotted
grey), and the IMF for the young Trapezium cluster (dashed grey)
\citep{Muench2002} already present in Figure~\ref{fig:2}. All these
stellar IMFs were shifted to higher masses by an average scale factor
of about 3.  The point of this comparison is to simply illustrate the
overall similarity between different stellar IMFs from different
environments and constructed in different ways, and the DCMF of the
Pipe Nebula.  Considering the overall uncertainties, apparent
differences between the different distributions are likely not
significant.  While Figure~\ref{fig:3} presents the background
subtracted core sample, the non background subtracted sample is
virtually identical in shape, although shifted towards higher masses.

\section{Discussion and Conclusions}

The close similarity in shape between the DCMF and the IMF supports a
general concept of a 1-to-1 correspondence between the individual
dense cores and soon to be formed stars.   In this respect our
observations are consistent with and appear to confirm the results of
dust continuum surveys in other clouds. However, our study provides
the first robust evidence for a departure from a single power law in
the core mass function. We find that the location of the break
indicates that there is a factor of about 3 difference in mass scale
between the two distributions, which implies that a uniform SFE, that
we estimate to be 30\%$\pm$10\%, will likely characterize the star
formation in these dense cores, across the entire span of (stellar)
mass. This efficiency is very similar to the maximum estimated for
young embedded clusters \citep{Lada2003}.

It is well known that the generation of a powerful outflow is a
natural consequence of the formation of a protostar in a dense core
\citep{Lada1985}.  It has been long suspected that such outflows could
be very disruptive to surrounding material and could play an important
role in limiting the mass which can accrete onto a protostar
\citep{Shu1987,Shu1991}.  \cite{Matzner2000} have theoretically
investigated the disruption of dense cores by protostellar outflows
and have concluded that outflow disruption of individual cores can
produce SFEs in the range of 25--75\%.  Moreover they found that SFE
is independent of stellar mass, therefore the stellar and core mass
functions should be of similar overall functional form. Recently,
\citet{Shu2004} have developed a self-consistent analytical theory for
the origin of the stellar IMF incorporating the feedback from
outflows.  In their picture, cores are formed from magnetized
turbulent gas and are disrupted by outflows with a resulting SFE
predicted to be $\approx$ 30\%.  The close correspondence of this
prediction with our estimated SFE for the Pipe cores may implicate
outflows as the key mechanism for core disruption and for setting the
final masses of the protostellar objects. \citet{Shu2004} also
predicted both a slope for the power-law portion of the core mass
function and a location of the break from the power-law form that are
very similar to those we measure in the Pipe Nebula.  

A number of other theoretical calculations have also produced DCMFs
similar in shape to that found here
(\citet{Padoan2002,Bonnell2006,Elmegreen2006}).  In particular,
numerical simulations of turbulent clouds predict DCMF shapes similar
to the stellar IMF. However it is difficult to compare these
calculations in more detail to the derived Pipe DCMF because, unlike
the \citet{Shu2004} model, the mass scales for these DCMFs are
presently unconstrained and appear to be dependent on model
parameters, such as the numerical resolution of the
simulations as well as the Mach number of turbulence which are difficult
to quantify observationally (e.g.,
\cite{Klessen2001,Ballesteros-Paredes2002,Gammie2003}).  Nonetheless we
expect that with appropriate modification of model parameters such
models could also be made consistent with the observations.

These considerations suggest that apart from a uniform efficiency
  factor, which is likely fixed by the generation of an outflow, the
  birth mass of a star could to be completely predetermined by the
  mass of the dense core in which it is born.


The existence of a characteristic mass in the stellar IMF does
suggests a characteristic mass scale is produced by the physical
process of star formation.  The physical origin of this mass scale has
not been clear although it has been long known that thermal (Jeans)
fragmentation can produce a mass scale for star formation
\citep{Larson1985,Larson1986}.  For a pressure bounded core the
appropriate mass scale is the critical Bonnor-Ebert mass,

\begin{equation}
  \label{eq:1}
M_{BE} \approx 1.15\times(n/10^4)^{-1/2}\times(T/10)^{3/2}
\end{equation}

\noindent
where $T$ is the gas temperature and $n$ its volume density.  The
characteristic mass scale suggested by the stellar IMF is about 0.3
\msun.  For a typical dark cloud of temperature of 10 K, the
corresponding density scale would need to be about $1.5 \times 10^5$
cm$^{-3}$.  This is significantly higher than the measured average
mean density of the Pipe cores of 8$\times$10$^3$ cm$^{-3}$.  However,
if the characteristic stellar mass scale is set by the characteristic
mass of the DCMF, which is a factor of $\sim$ 3.3 higher than that of
the stellar IMF, then the corresponding critical density is $\sim$
$1.3 \times 10^4$ cm$^{-3}$ much closer to that observed for the Pipe
dense cores.  This correspondence suggests that the fragmentation
scale for the DCMF and ultimately the stellar IMF may be set by a
process of thermal (Jeans-like) fragmentation, perhaps modified
somewhat by ambipolar diffusion and/or turbulence.  Interestingly,
from the recent analysis of Spitzer observations of the spatial
distribution of protostellar objects in NGC 2264, \citet{Teixeira2006}
found evidence for Jeans like fragmentation to be occurring in that
star forming region.

We note that the mean densities of the cores in our study range
  between 5$\times$10$^3$ and 2$\times$10$^4$ cm$^{-3}$ (for the
  background subtracted case).  This range is lower than that
  (3$\times$10$^{4-5}$ cm$^{-3}$) usually associated with typical star
  forming cores and even the starless cores typically traced by dust
  continuum observations \citep{Motte1998,Johnstone2001,Kirk2005}.
  Because extinction observations are more sensitive to lower column
  densities we are able to probe lower mass and density regimes than
  those probed by current dust continuum studies.  If the Pipe Nebula
  cores are to form stars, the relatively low mean densities they
  currently possess may indicate that they are in a very early stage
  of development.  This notion is somewhat strengthened by the absence
  of significant star formation activity in the Pipe nebula complex.
  Apparently, the functional form of the stellar IMF may be fixed
  during or before the earliest stages of core evolution.

The results of this letter suggest that the distribution of stellar
birth masses over essentially the entire range of stellar mass is the
product of the dense core mass function and a uniform SFE of
30\%$\pm$10\%. Hence, the question of the origin of the IMF is the
question of the origin of the DCMF, itself a result of the physical
process of cloud fragmentation/coalescence. The mass scale
characterizing the DCMF may result from a simple process of thermal
(Jeans-like) fragmentation in the molecular gas.

\begin{acknowledgements}
CJL acknowledges support from NASA ORIGINS Grant NAG 5-13041.
This publication makes use of data products from the Two Micron All
Sky Survey, which is a joint project of the University of
Massachusetts and the Infrared Processing and Analysis
Center/California Institute of Technology, funded by the National
Aeronautics and Space Administration and the National Science
Foundation.
\end{acknowledgements}






   
  



\bibliographystyle{aa}
\bibliography{/Users/jalves/Documents/Util/zmy}

\end{document}